\documentclass[aps,preprint,superscriptaddress,showpacs]{revtex4}
\usepackage{revsymb}
\usepackage[dvips]{graphicx}
\usepackage{here}

\begin{document}
\title{Applications of recurrence quantified analysis to study the
dynamics of chaotic chemical reaction} 

\author{H. Castellini}
\email{hcaste@fceia.unr.edu.ar}
\affiliation{Dpto\@. de F\'{\i}sica, F\@.C\@.E\@.I\@.A\@., 
Pellegini 250,
2000 Rosario}

\author{L. Romanelli}
\email{lili@ungs.edu.ar} 
\affiliation{Instituto de 
Ciencias, Universidad de General
Sarmiento, Roca 850, 1663 San Miguel,
Argentina}

\begin{abstract}
Recurrence plot is a quite easy tool to be used in time series
analysis,in particular for measuring unstable periodic orbits
embedded in a chaotic dynamical system. Recurrence quantified analysis (RQA) 
is an advance tool that allows the study of intrinsic complexity of dynamical 
system with a set of few parameters.
We use RQA for measuring chaotic transitions of NADH chemical reaction 
and determine numerically its characteristic parameters such as: 
Correlation integral, 
information entropy, Maximal Lyapunov's exponent, etc\@. 
For this work we have developed command sets with performance better
than TISEAN package

\end{abstract}
\maketitle

\section{Introduction}

\subsection{Recurrence quantified analysis}
Recurrence Plot (RP) was initially introduced by Eckman {\em et al}\@. (1987) 
\cite{r1}
as a tool for analyzing experimental time series data, especially useful 
for finding hidden correlations in highly complicated data and determine the 
stationarity of the time series. This method allows the identification of system 
properties that cannot be observed by the linear and nonlinear usual approaches 
It is worth to mention the simplicity of the algorithms during numerical 
calculations too.  
A RP is an injective application of a single reconstructed trajectory to the 
boolean matrix space, each pair $y_i$, $y_j$ coming from the time series is 
related with a pair $(i,j)$, called recurrence points. 
Let us consider $N$ values of a time series given by

\[ \{x_0, \cdots, x_{N-1}\}, \] 

 with $N$ large enough in order to evaluate the embedding dimension by using 
the false nearest neighborhood \cite{r2}
($d \ge 2$) and the time delay ($\tau \ge 1$) 
by looking at the relative minimum in the mutual information \cite{r3} 
Following Takens' embedding theorem \cite{r4}, 
the dynamics can be appropriately represented 
by the phase space trajectory reconstructed by using the time delay vectors 

\[ y_i=(x_i, x_{i+\tau}, \cdots, x_{i+(N-1) \, \tau}). \]

and the recurrence matrix is:

\begin{equation}
R_{(i,j)}=\Theta(\delta_h-||y_i-y_j||_{\infty}) -\Theta(\delta_l-||y_i-y_j||_{\infty})
\label{eq:1}
\end{equation}

where $\Theta$ is the Heaviside function and the matrix is symmetric. 
That means a RP is built by comparing
all delayed vectors with each other. A dark dot is plotted, ($R_{(i,j)}=1 $) with 
integer coordinates $(i,j)$ when 
 $\delta_l \le ||y_i-y_j||_{\infty} \le \delta_h$, 
otherwise a white dot is plotted ($R_{(i,j)}=0 $).
The interval $[\delta_l, \delta_h]$ it is known as {\em threshold corridor}.
The choice of this interval is critical, too large produces a saturation 
of the RP including irrelevant points, and too narrow losses information.
Since up to now in the literature there are not a satisfactory solution, 
an educated guess should be appropriate. 
In this work we used $[\sigma/10^5, \sigma/10^2]$, where $\sigma$ is the 
standard deviation.
 
Webber {\em et al}\@. \cite{r5} in order to characterize and analyze recurrent plots 
introduced a set of quantifiers, which are collectively called 
{\em recurrence quantified analysis} (RQA). The first of this quantifiers 
is the {\em \% recurrence} (\%REC), defined as:

\begin{equation} \texttt{\%REC}=100 \, \frac{N_r} {N_t} \end{equation}

where $N_t=\textrm{dim}(R)$ (every possible points) 
and $N_r$ is {\em number of recursive points} given by:

\[ N_r= 2 \# \{ (i,j) / R_{(i,j)} > 0  \, \textrm{and} \, i<j \}\]

The slope of the linear region in the S-shaped \%REC vs\@. corridor width 
is the correlation dimension. The second RQA quantifier 
is called {\em \% determinism} (\%DET); and it is defined as:

\begin{equation} \texttt{\%DET}=100 \, \frac{N_l}{N_r}  \end{equation}

where $N_l$ is called the {\em number of periodic dots} given by:

\[ N_l = 2 \# \{ (i,j) /  (i,j) \in d_c(k,b), \, i<j, 
\forall \, c, \, k, \, b>0 \} \]

and a periodic line with length $b$, origin $k$ and zone $c>0$ is defined as:

\[ d_c(k,b) = \{ (i,i+c) /  \prod_{i=k}^{k+b} R_{(i,i+c)} >0 \} \]

 The \%DET is related with the organization of the RP. 
The third RQA quantifier, called {\em entropy} (S), is closely related to \%DET. 

\begin{equation} S=-\sum^H_{b=1} P_b \log_2(P_b) \end{equation}

where $H$ is the length of the maximum periodic line, 
$P_n \ne 0$ is the relative frequency of the periodic lines with length $b > 0$.
The label entropy is just that, a label, not to be confused with Shannon's 
entropy since there is not a one to one correspondence between this quantifier 
and the Shannon's entropy. This quantity should be labeled more properly as 
{\em first rate cumulant} since is related with the relative frequency 
fluctuations. 
Moreover, for periodic orbits they are mapped onto diagonals with 
different lengths and uniformly distributed, giving values of $S>0$, with a maximum 
value. 
\label{t:1} Webber assumes that $S$ is related with 
Shannon's entropy if and only if the system is chaotic and the embedding dimension 
large enough.
The fourth quantifier is the longest periodic line found during the computation 
of \%DET given by $L_\textrm{max}$.  
Eckman {\em et al}\@. claim that line lengths on RP are directly related to 
inverse of the largest positive Lyapunov exponent. 
Short lines values are therefore indicative of chaotic or stochastic behavior.

\subsection{Dynamical system, a description}

The peroxidas-oxidase (PO) reaction is an important bridge between the chemical 
excitable oscillators Beluzov Zabotinsky reaction and biological oscillators
such as intracellular $\textrm{Ca}^{2+}$ oscillators \cite{r6}. 
It is now clear that PO reactions shows a wide spectrum of interesting 
behaviors including simple oscillations, bistability, quasiperiodicity and 
chaos \cite{r7}. 
These behavior have all been observed {\em in vitro} under well controlled 
laboratory conditions. This reaction appears in plants as part of the process of 
lignifications \cite{r8} with nicotinamide adenine dinucleotide (NADH) as electron 
donor. Its estequeometric formulae is:

\begin{displaymath}
2 \, \mathtt{NADH} + \mathtt{O}_2 + 2 \, \mathtt{H} \to 2 \,
\mathtt{NAD}^{+} + 2 \, \mathtt{H}_2\mathtt{O}
\end{displaymath}

In 1983, a model of PO reaction, now commonly referred to  
the \textbf{Olsen's Model} \cite{r9}, was proposed. Simulations with the Olsen model 
quantitatively reproduce both
the simple and chaotic oscillations of PO reactions. Studies with this model
showed that increasing the parameter $k_3$ caused the system to undergo 
a transition from simple oscillations to chaos via a cascade of periodic 
doubling bifurcations. The Olsen's model involves four variable, 
molecular oxygen \textbf{A}, 
NADH \textbf{B} and two intermediate species. One of the intermediates is very likely
$\textrm{NAD}^{+}$ \textbf{X}, while other is oxyferrous peroxidase \textbf{Y}, 
also know as \emph{compound III}. 
The complete mechanism corresponds system of four differential equations 
are given below:

\begin{eqnarray}
\frac{d A}{d t} & = & k_7-k_{-7} A-k_3 A B Y  \\
\frac{d B}{d t} & = & k_8-k_1 B X-k_3 A B Y  \\
\frac{d X}{d t} & = & k_1 B X-2 k_2 X^2 +3 k_3 A B Y -k_4 X +k_6  \\
\frac{d Y}{d t} & = & 2 k_2 X^2-k_5 Y -k_3 A B Y  
\end{eqnarray}

We use $k_3$ as control parameter which induces a transition to chaos 
type I \cite{r10}.

\section{Experiments and results}

\subsection{Experimental Setup}
 
If we wisht to use the RQA method to a time series long enough 
for reliability, we face with the usual difficulties, memory capacity and 
calculus complexity. 
Most of the codes were developed in C under Linux operating system.
This language does not implement the Boolean data type using integer type instead, 
so the required memory amount to store the recurrence matrix coming from a typical 
time series of $2^{16}$ data points is  128 Gbits and, 
the algorithm  complexity is quadratic, which implies huge time consuming. 
In order to solve these problems we propose to reduce the dimension of the 
matrix $R$, but then the mapping may not be one to one and a whole region of 
the reconstructed space would have the same points as an image.
Therefore $R$ as defined in (\ref{eq:1}) is modified as: $R_{(k,l)} > 0$
if and only if at least a pair of indexes exists

\[ (i,j) \in \{(i,j) / k=[i \frac{N}{m}], \,  l=[j \frac{N}{m}] \} \]

for $k$ y $l$ given, with $N=\textrm{dim}(R)$ and $m$ the amount data points, 
such as  

 \[ \Theta(\delta_h-||y_i-y_j||_{\infty}) - 
\Theta(\delta_l-||y_i-y_j||_{\infty}) >0 \]

and zero otherwise.

Due to hardware limitation we use recurrence matrices 10240 x 10240 values, although 
this fact was discussed before,and means a loss in the matrix resolution, the 
obtained results are quite satisfactory taking into account an adjusting of 
the corridor width and studying the parameters asymptotic behavior as a function 
of the number of the data.

\begin{figure}[H]
\begin{center}
\includegraphics[width=8cm, height=10cm, angle=-90]{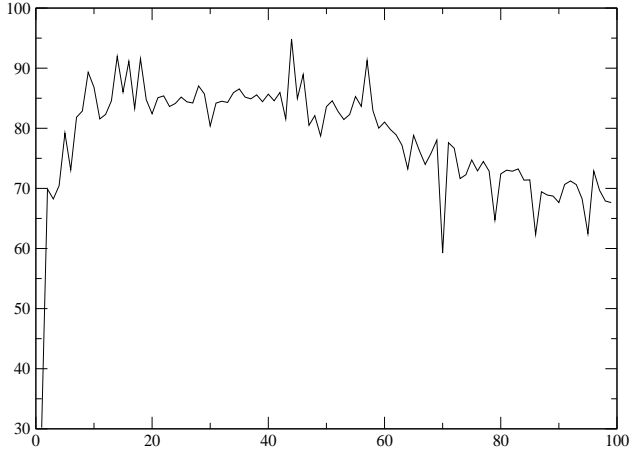}
\end{center}
\caption{\%DET as a function of cycles number for PO chemical reaction. 
For $k_3=0.028$ a limit cycle periodic behavior is observed}
\label{fig:1}
\end{figure}

Another problem concers the RQA high sensibility with the number of 
cycles during a time interval. 
As is depicted in figure-\ref{fig:1} \%DET has an asymptotic value when the 
number of cycles is greater than 30. 
If we are dealing with a sub-harmonic cascade or in the intermittency region an
abrupt change in the number of cycles is observed, giving us false transitions 
associate to the change in the cycles number.
But the number of cycles can not be so big because produces a graphics saturation 
then \%DET falls on minor values. A reason is an increment in vectors amount
mapped to same $(k,l)$ matrix component. So when this amount goes too far a
heuristic limit; determinism is lost. 
Taking this into account the parametric measurements associate to the RQA 
were done in such a way to use the half cycles on the plateau.

\subsection{Application to PO reaction}

\begin{figure}[H]
\begin{center}
\includegraphics[width=15cm, height=12cm, angle=-90]{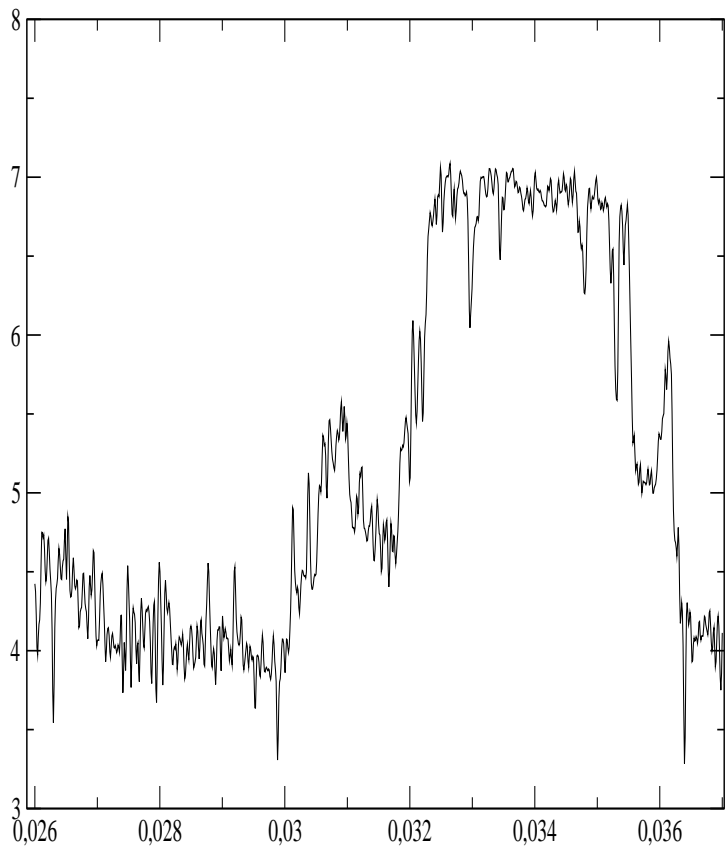}
\end{center}
\caption{$S(k_3)$ for 40 cycles, a plateau can be seen as in the chaotic 
as in the periodic regions.}
\label{fig:2}
\end{figure}

\begin{figure}[H]
\begin{center}
\includegraphics[width=8cm, height=10cm, angle=-90]{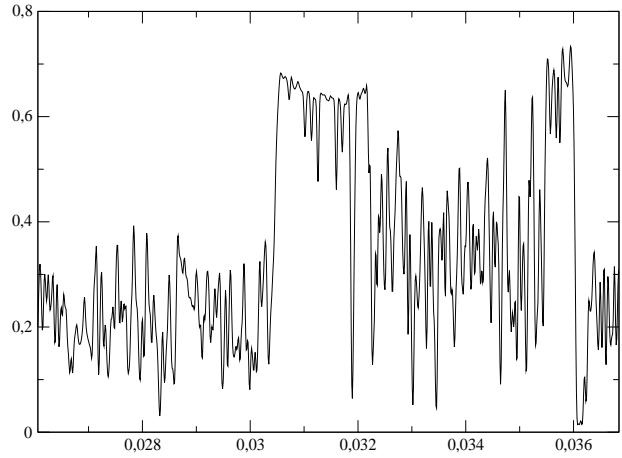}
\end{center}
\caption{Maximum Lyapunov exponent, $\lambda(k_3)$, 
found by using Aurell's {\em et al}\@. algorithm for 40 cycles.}
\label{fig:3}
\end{figure}

\begin{figure}[H]
\begin{center}
\includegraphics[width=8cm, height=10cm, angle=-90]{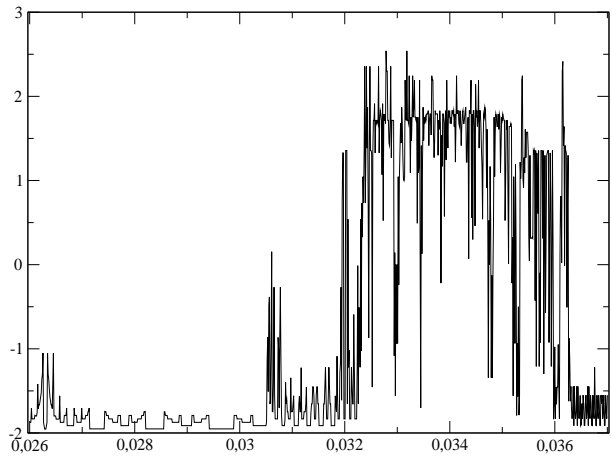}
\end{center}
\caption{$\gamma(k_3)=\log_2(L_{\textrm{max}})$ for 40 cycles.}
\label{fig:4}
\end{figure}

Since RP is not quite sensible to the election of the embedding dimension, $d$,
we use $d=9$ which this value allow us to eliminate the non diagonal points 
in the RP plots without affecting its general structure \cite{r11}.
On the other hand, no difference is observed neither $S(k_3)$,
or any RQA parameters within a range $3 \ge d \ge 15$, so we use the value 
which recommended by Takens' theorem \cite{r4}. 
As can be seen in figures-\ref{fig:2}, \ref{fig:3}, and \ref{fig:4} 
it was no found any direct proportion between
$S(k_3)$, and the maximum Lyapunov exponent, $\lambda(k_3)$,
by using the Aurell's algorithm {\em et al}\@. in TISEAN package \cite{r12}
calculated with a Thieler's window \cite{r13} with 500 time units.
Different algorithms associate with the evaluation of $\lambda(k_3)$ were 
used with unsatisfactory results due to the time consuming for getting a reliable 
slope in the graphical methods, so we developed our own software that allow us to 
apply the RQA methods for the calculation of the relevant parameters which 
characterize in any dynamical system independently of the size of the data.
It is important to note as depicted in figure-\ref{fig:2} the chaos transition 
characterization since $S(k_3)$ increase its value towards a plateau for the 
$k_3$ values which correspond to a sub-harmonic cascade, as well as a peak in 
the neighborhood of the first bifurcation value.

\begin{figure}[H]
\begin{center}
\includegraphics[width=8cm, height=10cm, angle=-90]{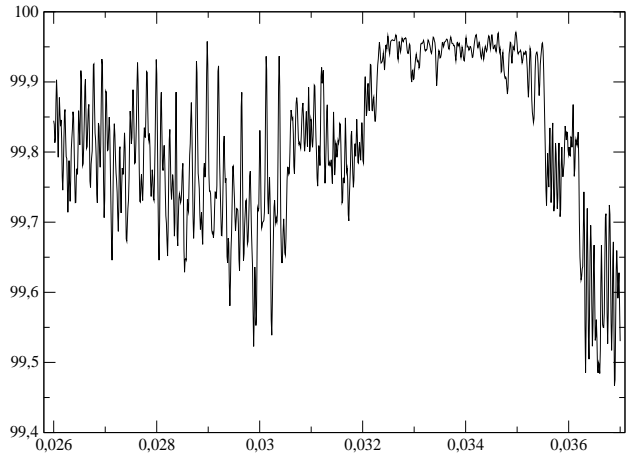}
\end{center}
\caption{\%DET as a function of $k_3$,it is not observed relevant changes 
in the curves shape due to the abundance of unstable periodic orbits.}
\label{fig:5}
\end{figure}

\%DET($k_3$) is associated with the existence of stable and unstable periodic orbits. 
In general the probability density related with periodic orbits is uniform, 
but for a dense family of unstable it orbits has an exponential distribution since 
it is more probable to find  our systems orbits with shorter period. 
For us, \%DET$\ge$99\%  always, see figure-\ref{fig:5}, and the behavior 
in the chaotic region is the same as $S(k_3)$, 
that means the increasing number of periodic orbits 
due to the sub-harmonic cascade from $k_3=0.031$ up to reach a maximum in the 
chaotic region densely populated by unstable orbits. For $k_3>0.0355$
the number of orbits is considerably reduced after an intermittent behavior. 
This is intrinsic to our dynamical system since it is not observed in the 
Lorentz system, used as a toy model \cite{r11}in this trail. 

On the other hand, by analyzing \%REC($k_3$) we may appreciate the scarce 
population of recurrent points in the RP. \%REC holds a constant value and is 
reduced at the beginning of the cascade with a minimum in the chaotic 
region. This behavior is different of that observed with \%DET and $S$ augmenting 
the value up to get a lower plateau. We assume this can be due to the changes 
in the shape of the attractor, since the same results is obtained for the Lorentz 
system \cite{r11}.

\begin{figure}[H]
\begin{center}
\includegraphics[width=8cm, height=10cm, angle=-90]{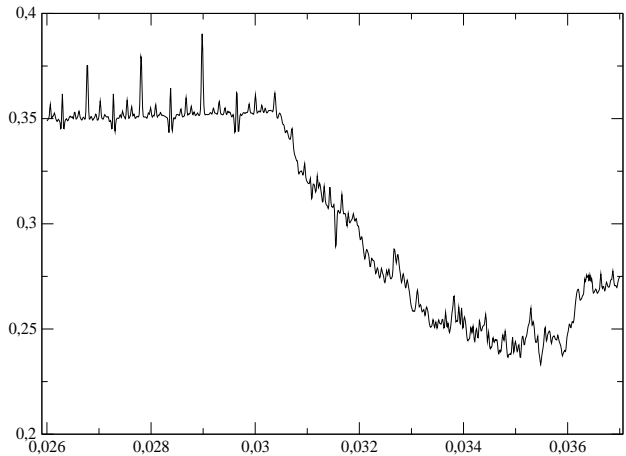}
\end{center}
\caption{\%REC as a function of $k_3$,it can be seen how scarcely populated is RP.}
\label{fig:6}
\end{figure}

By a simple inspection in the RP the increasing number of periodic orbits is clear, 
because more parallel lines to the diagonal appear, but the changes in the attractor 
shape are less impressive and only can be appreciate as alabeated curves when the 
embedding dimension is near $d=3$. 

\section{Conclusions}
In spite of Webber's assertion we do not observed any difference in the $S(k_3)$ 
graph neither for high embedding dimension, $d>9$, nor the lower ones, $2<d<10$. 
$S(k_3)$ shows a direct proportionality with the maximum Lyapunov exponent measured 
by other methods \cite{r14}. 
By the observation of the $\gamma(k_3)$, its behavior indicate the chaotic 
transition because is directly related with the maximum Lyapunov exponent 
besides an afin mapping. This is in agreement with Eckman {\em et al}\@. 
\cite{r1} statements.
Moreover, the algorithms developed and used in this work allow us a systematic 
RQA analysis efficiently.

\section{Acknowledgement}

We wish to thank the useful comments and discussions with Prof\@.
C\@. Webber which allow us to clarify issues regarding the RQA Method.
We also acknowledge the
``{\em Laboratiore de Physique des Milieux Ionises et Applications}'' 
and we wish to express our gratitude to Prof\@. G\'erard Bonhomme for 
providing his computer system.
This grant is partially supported by CONICET (PIP 4210).

\end{document}